\documentstyle[twocolumn]{mn}
\input epsf.sty
\begin{document}
\title{Correlation of Excursion Sets for Non-Gaussian CMB Temperature
Distributions}
\author[R.B.Barreiro, J.L. Sanz, E. Mart\'\i nez-Gonz\'alez and J. Silk]
{R.B.~Barreiro$^{1,2}$, J.L.~Sanz$^1$, E.~Mart\'\i nez-Gonz\'alez$^1$ 
and J.~Silk$^3$\\
$^1$ Instituto de F\'\i sica de Cantabria (CSIC-UC)\\ 
Facultad de Ciencias, Av. de Los Castros s/n, Santander 39005, SPAIN\\ 
$^2$ Departamento de F\'\i sica Moderna, Universidad de Cantabria\\ 
Facultad de Ciencias, Av. de Los Castros s/n, Santander 39005, SPAIN\\ 
$^3$ Astronomy Department and Center for Particle Astrophysics\\
University of California, Berkeley, CA 94720\\}
 
\maketitle
\begin{abstract}

We present a method, based on the correlation 
function of excursion sets above a given threshold, to test the Gaussianity 
of the CMB temperature fluctuations in the sky. In particular, this method
can be applied to discriminate 
between standard inflationary scenarios and those 
producing non-Gaussianity such as topological defects.
We have obtained the normalized correlation of excursion sets, including 
different levels of noise, for 2-point probability density functions 
constructed from the Gaussian, 
$\chi_n^2$ and Laplace 1-point probability density functions 
in two different ways.
Considering subdegree angular scales, we find that this method can 
distinguish between different distributions even if the corresponding marginal
probability density functions and/or the radiation power spectra are the same.

\end{abstract}

\begin{keywords}
cosmology: cosmic microwave background - anisotropies - Gaussianity
\end{keywords} 
 
\section{Introduction}

The temperature field $T(\theta, \phi)$, associated with the cosmic 
microwave background
(CMB), is usually assumed to be a homogeneous and isotropic Gaussian 
random field, i.e. the n-point 
probability density function is a multivariate Gaussian characterized 
by the 2-point 
correlation function $C(\theta )$ or equivalently the radiation 
power spectrum
$C_\ell$. The form of the temperature field at recombination is related 
to the initial 
density field coming from quantum fluctuations during an inflationary 
era in the very early stages of the history of the Universe.
However, other sources of density fluctuations  (such as
topological defects) 
may emerge at these early
times and generate a non-Gaussian random field.
In any case future high precision maps should be searched for any traces of
non-Gaussianity since both foregrounds and systematic errors can leave 
non-Gaussian imprints as well.

Whereas very accurate temperature maps can be obtained for the standard
inflationary model (at the level of $\la 1\%$ in the radiation power
spectrum), this is not the case for scenarios based on topological defects. In
particular, the best known of these models is the one generated by cosmic
strings (Turok 1996), which results in   temperature maps 
approximately Gaussian on scales
that span  from
relative small (a few arcminutes according to Gott III et al. 1990) to large
angular scales. Monopoles and textures generate maps which 
are also Gaussian at
large angular scales but not at small scales (Shellard 1996). The texture
and monopole maps show an asymmetric distribution with a positive tail
decreasing more slowly than in the Gaussian case (Turok 1996).

Taking into account the present uncertainty in characterizing the temperature
distributions for some non-standard scenarios, in this paper we will explore
several non-Gaussian models. We  consider a $\chi_n^2$ model for the 
temperature distribution function that satisfies
the previous two properties,  asymmetry and an exponential tail, and a Laplace
model that represents a symmetric distribution  with exponential tails.
Moreover, we will also study a toy model which has the property of 
having a Gaussian
marginal distribution but a non-Gaussian 2-point probability density function, 
a property that appears in the
cosmic string maps (Gott III et al. 1990). As it is shown in figures 12a,b 
of that paper, the temperature 1-point probability density function of cosmic
strings is Gaussian on scales above a few arcminutes whereas the one of the
temperature gradient shows a clear non-Gaussian behaviour. On the other hand,
the patchy behaviour in a CMB map associated to such network can be represented
by the mentioned toy model (see \S2.2).

In the analysis of the 2D temperature maps, we will consider 
the correlation of the
excursion sets (or regions) above a certain threshold. 
For the 3D matter density field,  such correlations were introduced by Kaiser 
(1984) assuming a
Gaussian 2-point probability density function to study the 
correlation function of rich clusters. He found an amplification 
with respect to the correlation function
associated with the underlying matter fluctuations. This was applied
(Mart\'\i nez-Gonz\'alez \& Sanz 1988) to test models of biased structure 
formation in several
scenarios. Coles (1989) modeled the matter density field with a 1D Gaussian, 
log-normal and a $\chi_1^2$ 
distributions (the 2 non-Gaussian distributions were used 
in an attempt to allow 
for non-linear evolution) to study
the correlation function of both peaks and excursion sets (in this last case,
the Politzer \& Wise 1984 approximation was assumed). The excursion sets are
easier to identify in the maps than the peaks and they carry very similar
information at high thresholds. In fact, the number of maxima and
excursion sets above a threshold coincides asymptotically.

For the 2D temperature field, non-Gaussian statistics have been applied 
by Coles \&
Barrow (1987) to obtain the mean size and frecuency of occurrence of 
the excursion sets above a given level
for different random fields. Such properties depend only on the 
1-point probability density function. 
 On the other hand, Pompilio et al. (1995) used a multifractal 
analysis of the temperature scans 
to explore the possibility of distinguishing between Gaussian 
fluctuations and non-Gaussian ones 
produced by a network of cosmic strings. Finally, Kogut et al. (1995) 
used the peak correlation
function in the COBE-DMR 2-year sky maps to test for a class of 
non-Gaussian models. In the last paper, 
the temperature maps are generated using a spherical harmonic 
decomposition whose multipole
components $a_{lm}$ are drawn from Gaussian, log-normal 
and ${\chi }^2_n$ distributions. They find that the 2-point correlation of
peaks is a better discriminator between Gaussian and non-Gaussian temperature
fields than the genus and the 3-point correlation function. Kogut et al. (1996)
also carried out a similar analysis with the DMR 4-year sky maps.

In a previous paper, we studied some geometrical properties of the CMB maxima
above a given level (mean number, curvature and ellipticity) assuming Gaussian
temperature fluctuations (Barreiro et al. 1997), to discriminate among
standard CDM models.
In this paper, we will use the correlation of excursion sets 
to test the Gaussianity of the CMB
temperature fluctuations in the sky, including the presence of noise. 
An advantage of using this quantity is that it is very easy to identify all
the excursion sets (i.e. pixels above a fixed threshold) in temperature maps.
These correlations are amplified by a purely statistical effect 
with respect to those associated with the temperature field. 
They carry information on 
the 2-point probability density function, allowing us
to distinguish between two 
different distributions even if
the underlying 1-point probability density function is the same. 
The latter point will become clear with the 
use of some reference toy models. 
 
We discuss, in \S2, 
the different 2-point probability density functions generated 
in two different ways. 
\S3 is dedicated to calculating the
correlation of excursion sets for different models in the case of an 
ideal experiment whereas in 
\S4 we include noise. Finally, in \S5 and \S6, 
we give the main results and conclusions
of the paper, respectively.  

\section{Two point distribution functions}

We are interested in several Gaussian and non-Gaussian distributions that
could represent the temperature field produced in different cosmological
scenarios. Inflationary models will generate Gaussian random fields whereas
other models such as those derived from topological defects will in general
give rise to non-Gaussian ones. We shall introduce the
chi-squared probability density function as a simple distribution 
that asymptotically contains the
Gaussian one. It has already been used in different cosmological contexts, in
relation to the large scale structure (Coles 1989), to study topological
properties of the CMB (Coles \& Barrow 1987) and as a test of
Gaussianity in the {\it COBE}-DMR four-year sky maps (Kogut et al. 1995).

\begin{figure}
\vspace{5cm}
\caption{The Gaussian (solid), $\chi_1^2$ (short-dashed), 
$\chi_{30}^2$ (dotted), $\chi_{60}^2$ (long-dashed) and Laplace (dotted-short
dashed) 1-pdf's with zero mean and unit variance are shown.}         
\end{figure}

The aim of this paper is to calculate the 2-point
correlation of the excursion sets above a given threshold. In order to do
this, we need to know the corresponding 2-point probability density functions 
(hereinafter, 2-pdf's). The Gaussian, $\chi^2$ with n degrees
of freedom and Laplace 2-pdf's are obtained from the 1-pdf's (see figure 1)
in two different
ways: one will follow the "standard" procedure whereas the other represents a
simple "toy" ansatz (Berry 1973, Jones 1996). 

\subsection {Standard procedure}

a) Gaussian

The 2-pdf for a Gaussian field with zero mean, variance $\sigma^2$ and 
correlation $\tau$ is given by:

\begin{equation}
f_X(x_1,x_2)= \frac{1}{2\pi \sigma^2 \sqrt{1- \tau^2}} 
e^ {- \frac{x_1^2+x_2^2-2 \tau x_1x_2}{2 \sigma^2 (1- \tau^2)} }
\enspace .
\end {equation}
This distribution is symmetric and extends from $-\infty$ to $\infty$.

\noindent
b) Chi-squared $\chi^2_n$

It is possible to generalize the univariate $\chi_n^2$ distribution with n
degrees of freedom to the bivariate case as follows: let us define the random
vector $y\equiv(y_1,y_2)=\sigma_x^2(x_1^2,x_2^2)$ where $\sigma_x$ is a
constant and $x\equiv(x_1,x_2)$ is a random variable described by a bivariate
Gaussian satisfying $\left<x_1^2\right>=\left<x_2^2\right>=1$, 
$\left<x_1x_2\right>=\tau_x$. Then, it is possible
to calculate the 2-pdf for y, as well as for the sum of n independent random
vectors of this type. The special case $\tau=0$ then corresponds to a pair of
independent $\chi_n^2$ univariate distributions with n degrees of freedom 
multiplied by $\sigma_x^2$. In the Appendix, we derive the 2-pdf for such a
bivariate $\chi_n^2$ distribution, resulting in the following expressions
($\sigma^2\equiv2n\sigma_x^4, \tau=\tau_x^2$):
 
$\tau<0$ :
\begin{eqnarray}
f_{\chi_n^2}(y_1,y_2) & = & \frac{n}{2\sigma^2(1-\tau)}
\left( \frac{-y_1y_2n}{2\tau\sigma^2} \right)^{\frac{n-2}{4}}
\frac{1}{\Gamma\left(\frac{n}{2}\right)} \nonumber\\
 & & e^{-\sqrt{\frac{n}{2}}
\frac{y_1+y_2}{\sigma(1-\tau)}} J_{\frac{n}{2}-1}\left(
\frac{\sqrt{-2n\tau y_1y_2}}{\sigma(1-\tau)}\right) \enspace ,
\end{eqnarray}
 
$\tau=0$ :
\begin{equation}
f_{\chi_n^2}(y_1,y_2) = \frac{1}{y_1y_2}\left(\frac{ny_1y_2}
{2\sigma^2}\right)^{\frac{n}{2}}
\left[\frac{1}{\Gamma\left(\frac{n}{2}\right)}\right]^2 
 e^{-\sqrt{\frac{n}{2}}
\frac{y_1+y_2}{\sigma}}  \, ,
\end{equation}
 
$\tau>0$ :
\begin{eqnarray}
f_{\chi_n^2}(y_1,y_2) & = & \frac{n}{2\sigma^2(1-\tau)}
\left( \frac{y_1y_2n}{2\tau\sigma^2} \right)^{\frac{n-2}{4}}
\frac{1}{\Gamma\left(\frac{n}{2}\right)} \nonumber\\
 & & e^{-\sqrt{\frac{n}{2}}
\frac{y_1+y_2}{\sigma(1-\tau)}} I_{\frac{n}{2}-1}\left(
\frac{\sqrt{2n\tau y_1y_2}}{\sigma(1-\tau)}\right) \enspace ,
\end{eqnarray}

\noindent where $J_\nu$ 
and $I_\nu$ are the Bessel and modified Bessel function 
of the first kind, respectively.

The $\chi_n^2$ distribution is asymmetric and extends from 0 to $\infty$.
Since the temperature field is assumed to have zero mean, we will center 
this distribution in the following calculations, which are also extended to 
negative values. 
In the limit $n \to \infty$, we recover a Gaussian process with mean 
$\mu$, variance $\sigma^2$ and correlation $\tau$.

\noindent
c) Laplace

The Laplace 1-pdf is given by:

\begin{equation}
f_Z(z)=\frac{1}{\sqrt2 \sigma} e^{\frac{-\sqrt2 \mid z\mid}{\sigma}}
\enspace .
\end{equation}

Following McGraw \& Wagner (1968), it is possible to generalize the Laplace
distribution to a 2-pdf if we assume elliptical symmetry properties. An
elliptically symmetric distribution is characterized by a 2-pdf whose contour
lines of equal height are ellipses in the $(z_1,z_2)$ plane and satisfies:
\begin {equation}
f(z_1,z_2,\tau)=f(R,\tau)
\enspace ,
\end {equation}
\noindent where $R=\sqrt{z_1^2+z_2^2-2z_1z_2\tau}$. 
The previous authors found for the Laplace 2-pdf:

\begin{equation}
f_Z(z_1,z_2)\!=\!\frac{1}{\pi\sigma^2\sqrt{1-\tau^2}} 
K_0 \left( \frac{\sqrt{2\left(z_1^2+z_2^2-2z_1z_2\tau\right)}}
{\sigma \sqrt{1-\tau^2}} \right)
\end{equation}

\noindent with mean value $\mu=0$, variance $\sigma^2$ and correlation
$\tau$. $K_0$ is the modified Bessel function of second kind.
This distribution is symmetric and extends from $-\infty$ to $\infty$.
Notice that for the case $\tau=0$ the Laplace 2-pdf is not given as the 
product of the 1-pdf's.

\subsection{Toy model procedure}

We will show in the present section a simple method to generalize a 1-pdf 
to a 2-pdf (Berry 1973, Jones 1996). This generalization has been used by the
first author to
describe the statistical properties of echoes diffracted from rough surfaces.
If we have a process with correlation $\tau$ and described by a 1-pdf 
f(x), with mean $\mu$ and variance
$\sigma^2$, then, the distribution:

\begin{equation}
f(x_1,x_2)=f(x_1)\delta (x_1-x_2) \tau +f(x_1)f(x_2)(1-\tau)
\end{equation}

\noindent is a 2-pdf with correlation $\tau$ and 
marginal distributions given by the original 1-pdf. Here $\delta(x-x_0)$ 
is the Dirac delta distribution centered 
on $x_0$. This distribution extends in the same range as the
f(x) being considered and satisfies the interesting properties:
$f(x_1,x_2)=f(x_1)\delta (x_1-x_2)$ for $\tau\rightarrow 1$ and
$f(x_1,x_2)=f(x_1)f(x_2)$ for $\tau \rightarrow 0$, i.e., the 
same limits that can be obtained for the bivariate Gaussian. 
This generalization is a simple example which shows that knowledge of  the 1-pdf and 
the correlation allows many possibilities for the 2-pdf.
We will construct this 2-pdf for the Gaussian and non-Gaussian fields
considered previously, and we will compare 
the correlation of regions obtained in this way
with those obtained from the standard 
distributions given in $\S2.1$. A CMB map associated to this toy model
represents a surface consisting of flat regions
separated by vertical walls whose rms height is $2\sigma$, $\sigma$ being the
dispersion of the underlying 1-pdf (Berry, 1973). In particular, for the
Gaussian 1-pdf this could mimic the patchy behaviour given by the
Kaiser-Stebbins effect produced by a network of long
straight strings (Kaiser \& Stebbins 1984, Pompilio et al. 1995). For a
simulation of the Kaiser-Stebbins effect see Magueijo \& Lewin (1997).

\section{Correlation of excursion sets}
Kaiser (1984) has calculated the 2-point correlation function of excursion 
sets above a threshold
$\nu \sigma $ for a 3D Gaussian field of variance $\sigma^2$ and 
mean value zero. We will follow the same procedure for the distributions
considered in this paper.

The probability that a randomly chosen point lies above a certain level $\nu
\sigma$ is:

\begin{equation}
P_1=\int_{\nu \sigma}^{\infty}f(x)dx
\enspace .
\end{equation}
If we choose another point at distance r from the first, the probability that
the field at both points takes a value exceeding that threshold is:

\begin{equation}
P_2=\int_{\nu \sigma}^{\infty} \int_{\nu \sigma}^{\infty}
f(x_1,x_2,\tau)dx_1dx_2
\enspace ,
\end{equation}

\noindent where $\tau$ is the correlation of the field.

Therefore, the 2-point correlation function for the excursion sets, 
$\xi_{>\nu}$, is:

\begin{equation}
1+\xi_{>\nu}(r)= \frac{P_2}{P_1^2}
\enspace .
\end{equation}
$\xi_{>\nu}(r)$ gives the fractional excess probability 
that a point $x_2$ 
lies above $\nu \sigma$, given that $x_1$ also exceeds that threshold 
and $\mid x_1-x_2 \mid =r$.

Next we shall introduce the normalized correlation $C_{>\nu}(r)$ 
associated to the characteristic function of an excursion set h(x) 
defined by:

\begin{eqnarray}
h(x) & = & 1 \qquad {\rm if~~~f(x)}>\nu \nonumber \\
h(x) & = & 0  \qquad {\rm otherwise.}
\end{eqnarray}
    
Then the correlation $C_{>\nu}(r)$ is the standard correlation 
coefficient of h(x): $C_{>\nu}(r) \equiv [\left<h(x_1)h(x_2)\right>- 
\left<h(x_1)\right>\left<h(x_2)\right>]/
[\sigma_h(x_1) \sigma_h(x_2)]$
with $\sigma_h^2(x) \equiv \left<h^2(x)\right>-\left<h(x)\right>^2$. 
It is straightforward to 
obtain the relation:

\begin{equation}
C_{>\nu}(r)= \frac {P_1} {1-P_1} \xi_{>\nu}(r) \quad .
\end{equation}

We note that $P_2(\tau=1)=P_1$ for the standard Gaussian, standard
chi-squared and toy model distributions. We have also proved that this result is
valid for high thresholds in the standard Laplace case, and it applies for the 
$\nu$'s considered in the present paper.

$\xi_{>\nu}(r)$ at zero lag ($\tau=1$) only contains information about the
1-pdf, however $C_{>\nu}(0)=1$ for any distribution. Since the information from
the 2-pdf is only encoded in the shape of the correlation of excursion sets and
not in its normalization, it is better to use $C_{>\nu}(r)$ to test the 
discriminating power of the 2-pdf.

The standard and toy model procedures will be considered separately.

\subsection {Standard procedure}

a) Gaussian

For a Gaussian field $P_1$ is given by:

\begin{equation}
P_1= \frac{1}{2} {\rm erfc}\left(\frac{\nu}{\sqrt2}\right)
\enspace ,
\end{equation}

\noindent where erfc(x)=$\frac{2}{\sqrt\pi}\int_x^\infty e^{-t^2} dt$   
is the complementary error function.

One of the integrals which appears in $P_2$ can also be evaluated, resulting: 

\begin{equation}
1+\xi_{>\nu}=
\frac{\left( 2/\pi \right)^{1/2}}
{\left[{\rm erfc}\frac{\nu}{\sqrt{2}}\right]^2}
\int_{\nu}^{\infty}dy e^{-\frac{y^2}{2}} 
{\rm erfc}\left( \frac{\nu-\tau y}{\sqrt{2\left( 1-\tau^2\right)}}\right)
\end{equation}
in agreement with Kaiser (1984).
This last expression must be evaluated numerically when estimating 
$C_{>\nu}$.

\noindent
b) Non-Gaussian

As explained in $\S2.1$, the chi-squared distribution has mean
value $\mu$, different from zero, and it must be centered if we want this
distribution to 
describe the temperature field. To take this into account, the
integrals appearing in $P_1$ and $P_2$ have been evaluated between
$\mu+\nu\sigma$ and $\infty$. 
This procedure is completely equivalent to centering the 
$\chi^2_n$ distribution and taking the limits as shown 
in equations (9) and (10). Then, $P_1$ is given by:

\begin{equation}
P_1=1- \frac{\gamma\left( \frac{n}{2}, \sqrt{\frac{n}{2}}\left( \nu+
\sqrt{\frac{n}{2}} \right)\right)}{\Gamma\left(\frac{n}{2} \right)}
\enspace ,
\end{equation}
\noindent where $\gamma(a,x)=\int_0^x e^{-t}t^{a-1} dt$ is 
the incomplete gamma function.

For the Laplace case, we have:

\begin{eqnarray}
P_1 & = & 1-\frac{1}{2}e^{\sqrt2 \nu}    \qquad \quad \nu<0 \enspace, \nonumber \\
P_1 & = & \frac{1}{2}e^{-\sqrt2 \nu}  \qquad \qquad \nu \geq 0 
\enspace .
\end{eqnarray}

For both the $\chi_n^2$ and Laplace distributions $P_2$, and therefore
$C_{>\nu}$, must be numerically evaluated.

\subsection {Toy model procedure}

For a 2-pdf generated as explained in $\S2.2$, $P_2=\tau P_1+ P_1^2(1-\tau)$ 
and so the correlation of regions has a very simple form:

\begin{equation}
C_{>\nu}(r)=\tau(r)
\enspace ,
\end{equation}

\noindent where $\tau$ is the correlation of the process.
 
\bigskip

\begin{figure}
\vspace{5cm}
\caption {We plot the correlation of regions above thresholds $\nu_s=2.5$ 
(upper
figures) and $\nu_s=3.5$ (lower figures) versus the correlation of the
cosmological signal, $\tau_s$, for $\Omega=1$, FWHM=10$'$, three
levels of noise (from left to right, $\sigma_N(10')=(0,1,3)\times 10^{-5}$) and
different standard distributions: Gaussian(solid line), 
$\chi_1^2$ (short-dashed), Coles
approximation for $\chi_1^2$ (dotted-long dashed, only in the plots with no 
noise), 
$\chi_{30}^2$(dotted), $\chi_{60}^2$ (long-dashed) and Laplace (dotted-short
dashed).}
\end{figure}

In the top left plot of figure 2, the normalized correlations of 
excursion sets above $\nu=2.5$
versus $\tau$ for a Gaussian, $\chi^2_n$ with n=1,30,60 degrees of freedom and 
Laplace standard distributions are shown. 
The different distributions go to zero when $\tau$ goes to zero, except for
the Laplace case due to the fact that its 2-pdf does not tend to the 
product of the 1-pdf's. 
We see that there are large quantitative differences between the Gaussian 
and non-Gaussian distributions. 
As expected, when n increases the $\chi_n^2$ correlation approaches  the
Gaussian one.
For comparison, the $C_{>\nu}$ for $\chi_1^2$ calculated 
by Coles (1989) using the
approximation of Politzer \& Wise (1984) is also plotted.
It is shown that this approximation only applies acceptably well
for small $\tau$'s, as  was already pointed out by Coles.
For high values of $\tau$ the approximation
clearly overestimates the actual normalized correlation of regions 
for the $\chi_1^2$ distribution.
In the bottom left corner of figure 2, the same distributions are 
plotted for a threshold $\nu=3.5$. 
We see that increasing the threshold amplifies 
the difference between the Gaussian and non-Gaussian distributions.
Notice that in the case of the toy model, $C_{>\nu}$ is given by the simple 
relation of equation (18).
We emphasize that even if we have two fields with the same correlation and
1-pdf (as the case for the standard and toy model procedure with the same
marginal pdf), we can discriminate between them using the correlation 
of regions.

\section{Correlation of excursion sets including noise}

In a real experiment, the presence of instrumental noise modifies the
correlations calculated in $\S3$. In addition, the result will depend on the
amplitude of the signal and then we need to choose specific cosmological
models. 
Whereas for the standard inflationary scenario with Gaussian temperature
fluctuations, the radiation power spectrum can be determined very accurately
once the matter content is specified, this is not the case for other
alternative scenarios such as topological defects. 
We will assume the same radiation power spectrum for all the Gaussian
and non-Gaussian distributions that we consider. We expect that the differences 
in a more realistic case, where the appropriate power spectra are known, 
are at least as large as the ones found under this assumption.
In particular, we consider in the present section a flat CDM model 
(baryon content $\Omega_b=0.05$,
Hubble constant h=0.5, cosmological constant $\Lambda=0$) with 
adiabatic fluctuations and a Harrison-Zel'dovich
primordial spectrum, kindly provided by N.Sugiyama. The
$C_\ell$'s have been normalized to the COBE 2-year maps (Cay\'on et al. 1996;
this normalization does not appreciably change with the 4-year data) being
the signal dispersion
$\sigma_s=3.5\times 10^{-5}$ for an antenna of FWHM=$10'$ and a smoothing
of the same width. We also include the effect of noise that is assumed to be 
Gaussian white noise.
Following standard observational procedures, we have filtered signal plus noise
with a Gaussian with the same width as the antenna.

The angular correlation function for the cosmological signal 
with a Gaussian beam profile is given by:

\begin{equation}
C(\alpha,\sigma_f)=\frac{1}{4\pi}\sum_\ell (2\ell+1) C_\ell
e^{-2\ell(\ell+1)\sigma_f^2}P_\ell(\cos\alpha)
\enspace ,
\end{equation}

\noindent and its normalized correlation $\tau_s$:
\begin{equation}
\tau_s=\frac{C(\alpha,\sigma_f)}{\sigma_s^2}
\enspace ,
\end{equation}
\begin{equation}
\sigma_s^2=C(0,\sigma_f)
\enspace ,
\end{equation}

\noindent where $\sigma$ is the Gaussian dispersion of the antenna
($\sigma_f=0.425 \times $FWHM).

The correlation $\tau_N$ and the dispersion $\sigma_N$ for the filtered
noise are very well approximated by ($\sigma_f \la 0.1$ rad):

\begin{equation}
\tau_N \simeq e^{-\frac{\alpha^2}{4\sigma_f^2}}
\enspace ,
\end{equation}

\begin{equation}
\sigma_N^2 \simeq \frac{A_N}{4\pi\sigma_f^2}
\enspace .
\end{equation}

We choose different levels of noise, fixing the noise amplitude 
$A_N=(0,1.9,17)\times 10^{-15}$ in order to obtain
$\sigma_N=(0,1,3)\times 10^{-5}$ after smoothing with a $10'$ FWHM Gaussian
window. These levels of noise cover the range of sensitivities expected for 
future experiments (e.g. MAP, Planck Surveyor Mission).

The presence of noise produces distortions in the distributions considered. The
new distribution is given by a sum of a Gaussian process (coming from the
noise) with dispersion $\sigma_N$ plus the process that characterizes the 
signal with dispersion $\sigma_s$. The new 1-pdf and 2-pdf are given by:

\begin{equation}
f(z)=\int f_N(z-y)f_s(y)dy
\enspace ,
\end{equation}

\begin{equation}
f(z_1,z_2)=\int\!\!\!\int f_N(z_1-y_1,z_2-y_2)f_s(y_1,y_2)dy_1,dy_2
\enspace ,
\end{equation}

\noindent where $f_N$ and $f_s$ refer to the distributions of the noise
and the signal respectively. The integrals are evaluated over the range covered by the given signal distribution.
The variance of the new process is given by $\sigma_t^2=\sigma_s^2+\sigma_N^2$ 
and its correlation, 
for angles greater than the coherence
angle $\theta_c^N $of the noise ($\theta_c^N \simeq
\sqrt2 \sigma_f$), can be approximated by the expression
$\tau_t \simeq \tau_s/(1+SNR^{-2})$  where 
$SNR=\sigma_s/\sigma_N$ is the signal--to--noise ratio.
We note that we are interested in temperature thresholds 
$\nu_s$ in units of $\sigma_s$, which is related to the threshold
$\nu_t$, in units of $\sigma_t$, by
$\nu_t=\nu_s/\sqrt{1+SNR^{-2}}$.

We will again consider both procedures separately.

\subsection{Standard procedure}

a) Gaussian         

For a signal described by a Gaussian distribution, the new process remains
Gaussian with dispersion $\sigma_t$ and correlation $\tau_t$ defined
as before.
Therefore, the functional form of the correlation of excursion sets is the same
as the one without noise given by equations (14) and (15), where 
the threshold and correlation are now $\nu_t$ and $\tau_t$, respectively.

\noindent
b) Non-Gaussian

$P_1$ and $P_2$ are again given by equations (9) and (10). For $P_1$ we have:

\begin{equation}
P_1=\frac{1}{2}\int f_s(y) {\rm erfc} \left[ g(y) \right] dy
\enspace ,
\end{equation}
 
\noindent where $g(y)=(\nu_t
\sqrt{1+SNR^2}+\left( \frac{\mu}{\sigma_s}-y \right)SNR)/\sqrt2$
with $\mu=\sqrt{\frac{n}{2}} \sigma_s$ and $\mu=0$ 
for the $\chi_n^2$ and 
Laplace cases, respectively. For the Laplace distribution $P_1$ can 
be evaluated analytically.

On the other hand, if we assume $\tau_N \ll 1$, some of the integrals which
appear in $P_2$ can
be evaluated, resulting in :
 
\begin{equation}
P_2 \!=\! \frac{1}{4}\int\!\!\! \int f_s(y_1,y_2) {\rm erfc} \left[ g(y_1)
\right]
{\rm erfc} \left[ g(y_2)\right]
dy_1 dy_2
\enspace .
\end{equation}

This approximation is valid for $\alpha \ga \theta_c^N$. 
To obtain a better estimation of the normalized correlation of regions for
small angles when noise is present, we have evaluated $P_2$ as a power series
considering terms up to second order in $\tau_N$ and then interpolated using
its known value at $\tau=1$ (again $P_2(\tau=1)=P_1$).

\subsection{Toy model procedure}

In this case $P_1$ and $P_2$ are also given by equations (26) and (27) where
$f_s(y)$ is the 1-pdf corresponding to the signal (Gaussian, $\chi_n^2$ or
Laplace) and $f_s(y_1,y_2)$ is the 2-pdf constructed using the toy model
procedure. In particular, assuming $\tau_N \ll 1$ we obtain for $P_2$:

\begin {equation}
P_2=\frac{\tau_s}{4}\int f_s(y) \left( {\rm erfc} \left[g(y)\right] 
\right)^2 dy + \left(1-\tau_s \right) P_1^2 \enspace .
\end {equation}

\bigskip

In figure 2, $C_{>\nu}$ versus $\tau_s$ are plotted for different
levels of noise and thresholds $\nu_s$ (for $\Omega=1$ and FWHM=$10'$), 
for the Gaussian, $\chi_n^2$ with n=1,30,60 and Laplace distributions following
the standard procedure. 
The main effect of the noise is to produce a rapid fall of the normalized
correlation of regions at small scales.
However, it is still
possible to see clear differences between Gaussian and non-Gaussian 
distributions, up to levels of SNR$\sim$1.
As before, increasing the threshold $\nu_s$ produces an amplification of the
differences.

\section{Results}

We have applied the calculations of the previous sections to open 
and flat CDM models
($\Omega=1,0.3,0.1$, with $\sigma_s(10')=3.5\times 10^{-5}$) and 
considered three different levels of noise
($\sigma_N(10')=(0,1,3)\times 10^{-5}$).

Besides the instrumental noise, our calculations are affected by
statistical errors with  whole sky coverage. Since we are considering very 
small angular scales, we expect that the cosmic variance will introduce a 
small uncertainty in our results. 

As we have already pointed out,
increasing the threshold $\nu_s$, enhances the difference between the considered
correlations. However, this also produces a rapid fall in the number of 
excursion sets above $\nu_s$ and, thus, higher statistical errors.
On the other hand, increasing the width of the antenna improves the SNR but 
drastically decreases the number of spots above $\nu_s$. This
shows that we must find a compromise for choosing the optimal 
parameters for our analysis.
In particular, we have taken an antenna FWHM=10$'$ and
a threshold $\nu_s=2.5$ to present our results.

\begin{figure}
\vspace{5cm}
\caption {In the four upper figures we plot the normalized correlation of 
regions for the
toy model (thicker line, labelled t) and standard (thinner line, labelled s) 
2-pdf's derived from the
same 1-pdf. In the lower figures comparisons between correlation of regions for
Gaussian and non-Gaussian
distributions derived from standard 2-pdf's are shown. In all the plots, we
have taken $\nu_s=2.5$, $\Omega=1$, FWHM=10$'$ and no noise. The different
lines
correspond, in all the cases, to the correlation of regions derived 
from Gaussian (solid), $\chi_1^2$ (short-dashed),
$\chi_{30}^2$ (dotted), $\chi_{60}^2$ (long-dashed) and Laplace (dotted-short
dashed) distribution.}  
\end{figure}

The normalized correlation of excursion sets above a threshold gives 
us different
information from that of the 2-point correlation of the temperature
fluctuations, since we obtain additional information about the 2-pdf. For this 
reason, our method is of especial interest for discriminating 
between different distributions of the temperature fluctuations for a given
power spectrum. 
In fact, as we have already pointed out, even if we have two fields 
described by the same 1-pdf
and correlation, we can show that it is possible
to discriminate between them through the correlation of regions.

In the upper part of figure 3, we plot pairs of standard and 
toy model distributions derived
from the same 1-pdf for $\Omega=1$, FWHM=10$'$, $\nu_s=2.5$ and no noise. 
The drawn errors have been estimated for all the distributions from high
resolution Gaussian simulations of the sky. We expect them to be a good
approximation to the real errors for the $\chi_{30}^2$ and
$\chi_{60}^2$ distributions. In the rest of the cases, even if the errors were
underestimated, we don't expect our results to be very much affected since 
the difference between distributions is, generally, very clear.
In addition, performing a simple Poissonian calculation of errors 
suggests that they are larger in the Gaussian case.
Using a $\chi^2$ test, we obtain a null result for the hypothesis
that each pair of curves is derived from the same population, 
with a very high confidence level ($\ga 99\%$). 
Considering open models, $\Omega=0.3,0.1$, or introducing
noise up to $\sigma_N(10')=3\times10^{-5}$ 
does not significantly affect these results.

On the other hand, we can compare the Gaussian and non-Gaussian standard
distributions. For $\Omega=1$, FWHM=10$'$, $\nu_s=2.5$ and no noise (see lower
plots of figure
3), we can distinguish between the Gaussian and non-Gaussian ($\chi_n^2$, 
n=1,30,60 and Laplace) standard distributions with
a confidence level $\ga 99\%$. 
Increasing the level of noise up to $\sigma_N(10')=3\times10^{-5}$ or 
changing the power spectrum does not appreciably modify these results.

\begin{figure}
\vspace{5cm}
\caption {We show the correlation of regions for flat (solid line) and 
open models
with $\Omega=0.1$ (dashed line, in upper figures) and $\Omega=0.3$ (dotted
line,
in lower figures) for different standard 2-pdf's (from left to right, Gaussian,
$\chi_{30}^2$ and Laplace). For all the figures, we have considered
$\nu_s=2.5$, FWHM=10$'$ and no noise.}
\end{figure}
                             
We can also try to discriminate between flat and open models for a given 
standard distribution through $C_{>\nu}$. 
For $\nu_s=2.5$, FWHM=10$'$ and no noise, we can differentiate between 
the flat and open ($\Omega=0.1$ or $\Omega=0.3$) models 
with a confidence level $\sim 99\%$.
In Figure 4, we have represented the Gaussian, $\chi_{30}^2$ and Laplace
distributions for the three considered values of $\Omega$, $\nu_s=2.5$, 
FWHM=10$'$ and no noise.

Considering a higher threshold amplify the difference among the
considered distributions. However it also produces a rapid fall in the number
of excursion sets above $\nu$ and thus higher statistical errors, what leads to
smaller confidence levels in the separation of the models.
On the other hand, the smaller the threshold the
closer $C_{>\nu}$ for the different distributions, since in the 
limit $\nu\to-\infty$
all the pixels are above $\nu$ and the correlation of regions 
is zero for all the distributions.
\section{Conclusions}

We have presented a method for testing the Gaussianity of the CMB temperature
fluctuations in the sky based on the normalized correlation function of
excursion sets above a given threshold. It can be used to distinguish between
standard inflationary scenarios of generation of density fluctuations and
those based on topological defects as well as to search for any trace of
non-Gaussianity due to systematic errors or foregrounds.
Such a technique was first introduced in
cosmology by Kaiser (1984) to study the correlation of galaxy clusters in
the context of biased scenarios of galaxy formation.

As an application, we have constructed 2-pdf's from the Gaussian, 
$\chi_n^2$ and Laplace 1-pdf's in two different ways.
From these, we have obtained the normalized correlation of excursion sets 
including different levels of noise. This correlation
contains additional information to that of the simple radiation power 
spectrum. 

Our main conclusion is that, using the correlation of excursion sets above
high thresholds (e.g. $\nu_s=2,3$) on subdegree scales, it is 
possible to discriminate between different distributions even if the 1-pdf 
and correlation are
the same. In particular, we are able to clearly distinguish between the
Gaussian and $\chi_{60}^2$ cases even in the presence of certain levels of 
noise within the range of sensitivities expected for future experiments.
Increasing the threshold amplify 
the differences among the considered distributions but also the statistical
errors, leading to smaller confidence levels.

Finally, the correlation of regions can also be used to compare different 
$\Omega$ models though it is less efficient than directly using the radiation
power spectrum.

\section*{Acknowledgements}

We would like to thank N. Sugiyama for providing us with  the radiation 
power spectrum and B. Jones for suggesting  a method of generalizing any 1-pdf
to a 2-pdf.
EMG, JLS and RBB acknowledge financial support from the Spanish DGES, 
project PB95-1132-C02-02, CICYT, project ESP96-2798-E and from Comisi\'on Mixta Caja Cantabria-Universidad
de Cantabria. RBB acknowledges a Spanish M.E.C. Ph.D. scholarship. 
RBB, JLS and EMG thank the NSF Center for Particle Astrophysics in Berkeley
for its hospitality during their stay there.

\appendix
\section{The 2-pdf for the $\chi_n^2$ distribution}

We can find the 2-pdf for $\chi_n^2$ in the following way. First, we construct
the 2-pdf for $\chi^2_1=X^2$, where X is a bivariate Gaussian process with 
zero mean. If $f_X(x_1,x_2)$ denotes its 2-pdf then:

\begin{equation}
f_{\chi^2_1}(y_1,y_2)=\mid J \mid f_X \left(x_1=x_1(y_1),x_2=x_2(y_2) \right)
\enspace ,
\end{equation}

\noindent being $\mid J \mid$ the Jacobian of the transformation 
$(x_1,x_2) \to (y_1,y_2)$, with $y_1=x_1^2$, $y_2=x_2^2$:

\begin{eqnarray}
f_{\chi_1^2} (y_1,y_2) & = & \frac{1}{2\pi \sigma_x^2 \sqrt{1- \tau_x^2} }
\frac {1} {\sqrt{y_1y_2}} e^{-\frac{y_1+y_2}{2\sigma_x^2(1-\tau_x^2)}}
\nonumber\\
 & & \cosh\left( \frac{\tau_x \sqrt{y_1y_2}} {\sigma_x^2 (1-\tau_x^2)} \right)
\enspace ,
\end{eqnarray}

\noindent where $\sigma_x^2$ and $\tau_x$ are the variance and correlation of 
the Gaussian process, respectively.

Taking into account that the characteristic function of a given distribution is
defined as the Fourier transform:

\begin{equation}
\phi(t_1,t_2)=\int \limits_{-\infty}^{+\infty} 
\int \limits_{-\infty}^{+\infty} dy_1 dy_2 f(y_1,y_2)e^{i(t_1y_1+t_2y_2)}
\enspace ,
\end{equation}

\noindent we get for $\chi_1^2$:

\begin{equation}
\phi_{\chi_1^2}(t_1,t_2)=\frac{1}{\sqrt{1-2i\sigma_x^2(t_1+t_2)
-4\sigma_x^4(1-\tau_x^2)t_1t_2}}
\enspace .
\end{equation}

On the other hand, the characteristic function of a sum of n independent
processes is the product of their characteristic functions, so we have
for $\chi_n^2$ :

\begin{equation}
\phi_{\chi_n^2}(t_1,t_2)=\left[ \phi_{\chi_1^2}(t_1,t_2) \right] ^n
\enspace .
\end{equation}
This distribution has variance $\sigma^2 \equiv 2n \sigma_x^4$, correlation
$\tau \equiv \tau_x^2$ and mean 
$\mu \equiv n\sigma_x^2 \equiv \sqrt { \frac{n}{2} }\sigma$.
Then, we can write $\phi_{\chi_n^2}$ in terms of $\sigma$ and $\tau$:

\begin{equation}
\phi_{\chi_n^2}(t_1,t_2)\!=\!\!\left[\frac{1}{1-i\sqrt{\frac{2}{n}}
\sigma(t_1+t_2)
-\frac{2}{n}\sigma^2(1-\tau)t_1t_2} \right]^{\frac{n}{2}}\!\!.
\end{equation}

Inverting eq.(A3) we obtain the 2-pdf for a chi-squared process with n
degrees of freedom given by equations (2-4) of $\S2.1$.


\begin{thebibliography}{}
\bibitem[]{} Barreiro, R.B., Sanz, J.L., Mart\'\i nez-Gonz\'alez, Cayon, L. \&
Silk, J. 1997, ApJ, 478, 1 
\bibitem[]{} Berry, M.V. 1973, Phil. Trans. R. Soc. (London), 273, 49
\bibitem[]{} Cay\'on, L., Mart\'\i nez-Gonz\'alez, E., Sanz, J.L., 
Sugiyama, N. \& Torres, S. 1996, MNRAS, 279, 1095 
\bibitem[]{} Coles, P. 1989, MNRAS, 238, 319
\bibitem[]{} Coles, P. \& Barrow, J.D. 1987, MNRAS, 228, 407
\bibitem[]{} Gott III, J.R., Park, C., Juszkiewicz, R., Bies, W.E., Bennett,
D.P., Bouchet, F.R. \& Stebbins, A. 1990, ApJ, 352, 1
\bibitem[]{} Jones, B. 1996, in The Cosmic Background Radiation, Proceedings of
the NATO ASI held in Strasbourg (France), Lineweaver et al. eds. in
press
\bibitem[]{} Kaiser, N. 1984, ApJL, 284, L9
\bibitem[]{} Kaiser, N. \& Stebbins, A. 1984, Nature, 310, 391
\bibitem[]{} Kogut, A., Banday, A.J., Bennett, C.L., G\'orski, K., Hinshaw, G.,
Smoot, G.F. \& Wright, E.L. 1996, ApJL, 464, L29                                
\bibitem[]{} Kogut, A., Banday, A.J., Bennett, C.L., Hinshaw, G., Lubin, P.M.
\& Smoot, G.F. 1995, ApJL, 439, L29 
\bibitem[]{} Magueijo, J. \& Lewin, A. 1997, astro-ph/9702131
\bibitem[]{} Mart\'\i nez-Gonz\'alez, E. \& Sanz, J.L. 1988, ApJ, 324, 653 
\bibitem[]{} McGraw, D.K. \& Wagner, J.F. 1968, IEEE Transactions on
information theory, Vol. IT-14, No.1 
\bibitem[]{} Politzer, H.D. \& Wise, M.B. 1984, ApJL, 285, L1
\bibitem[]{} Pompilio, M.P., Bouchet, F.R., Murante, G. \& Provenzale, A. 1995,
ApJ, 449, 1
\bibitem[]{} Shellard, P. 1996, in The Cosmic Background Radiation, Proceedings
of the NATO ASI held in Strasbourg (France), Lineweaver et al. eds. in press
\bibitem[]{} Turok, N. 1996, astro-ph/9606087
\end{thebibliography}
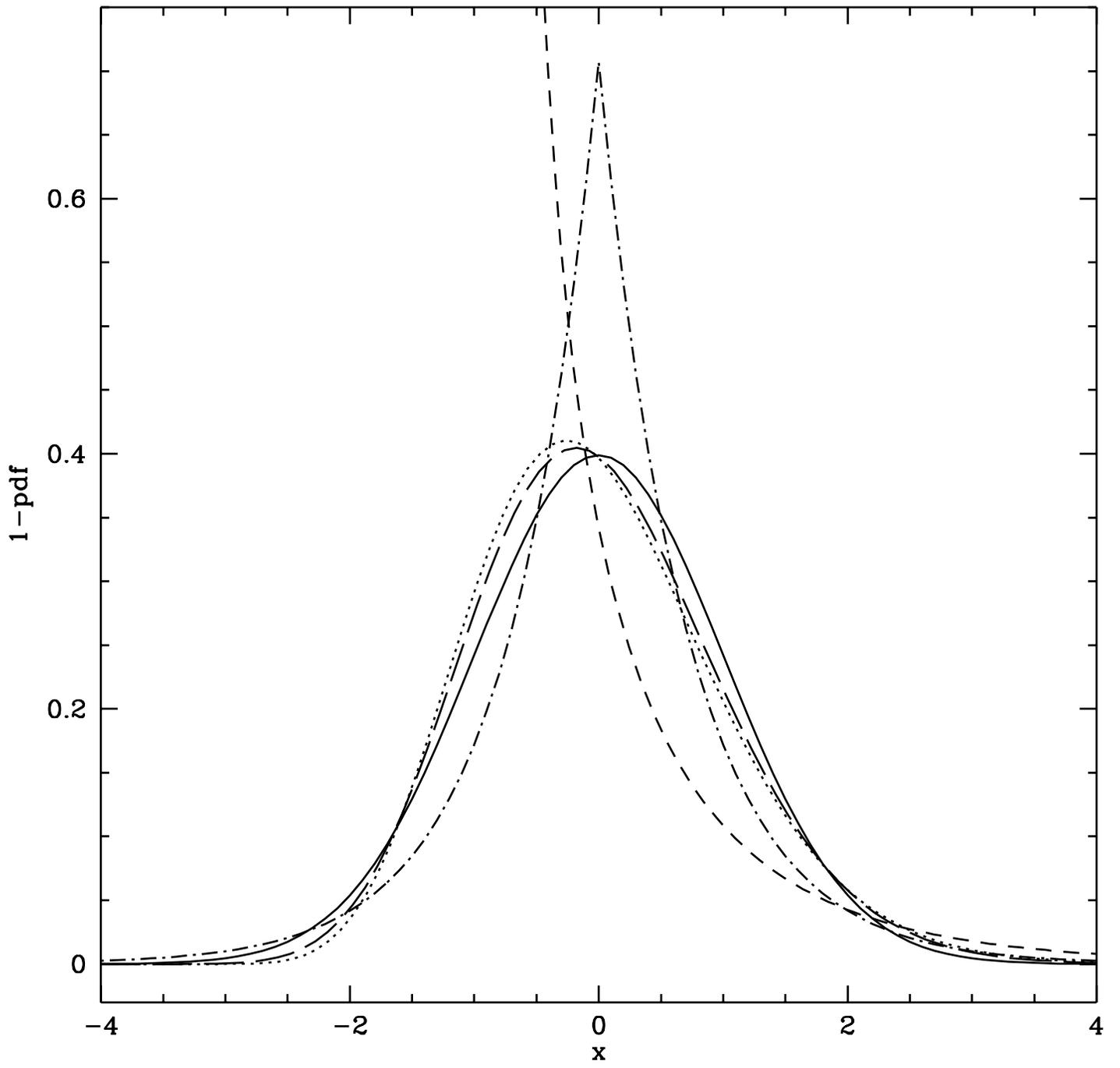
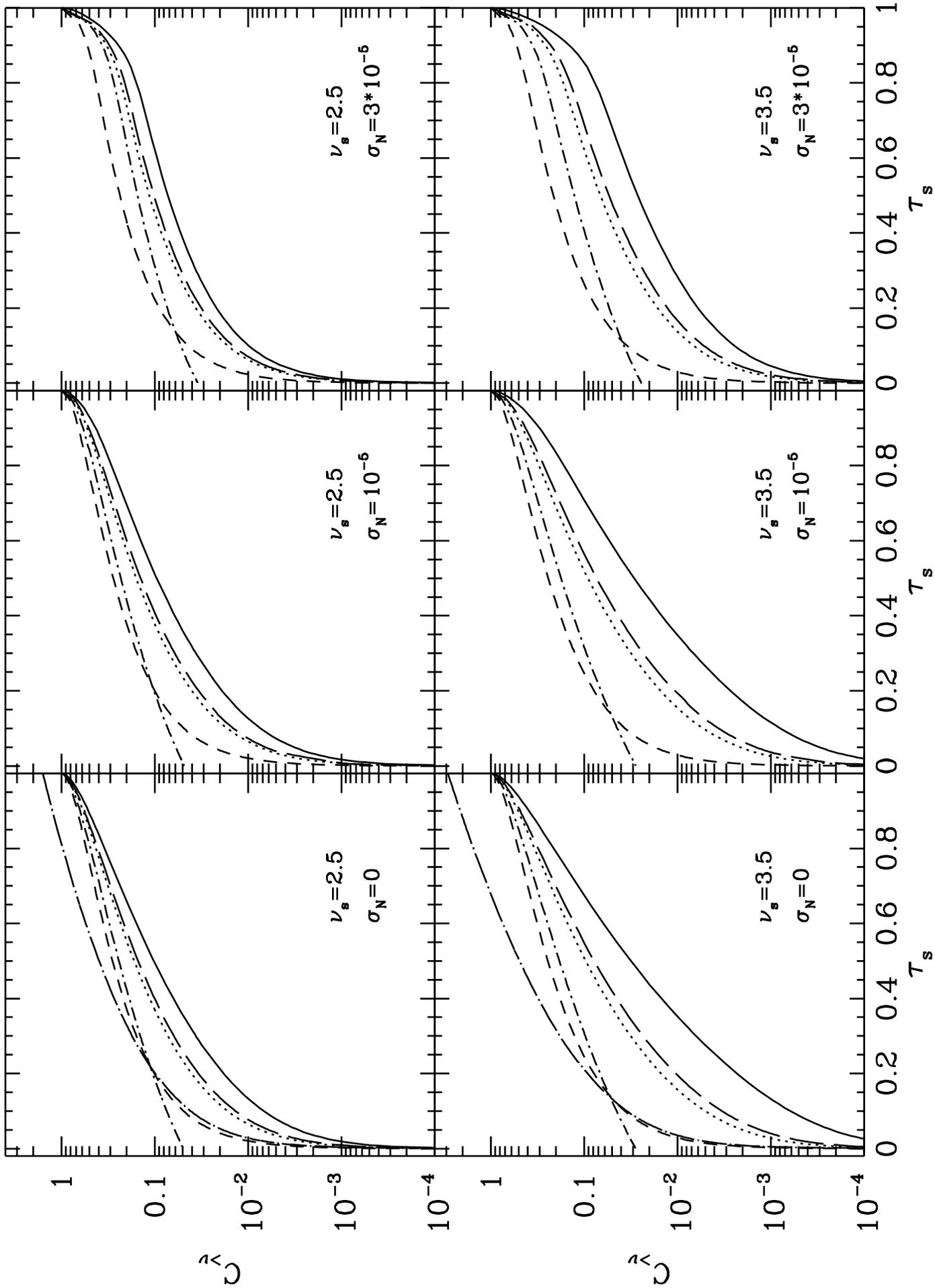
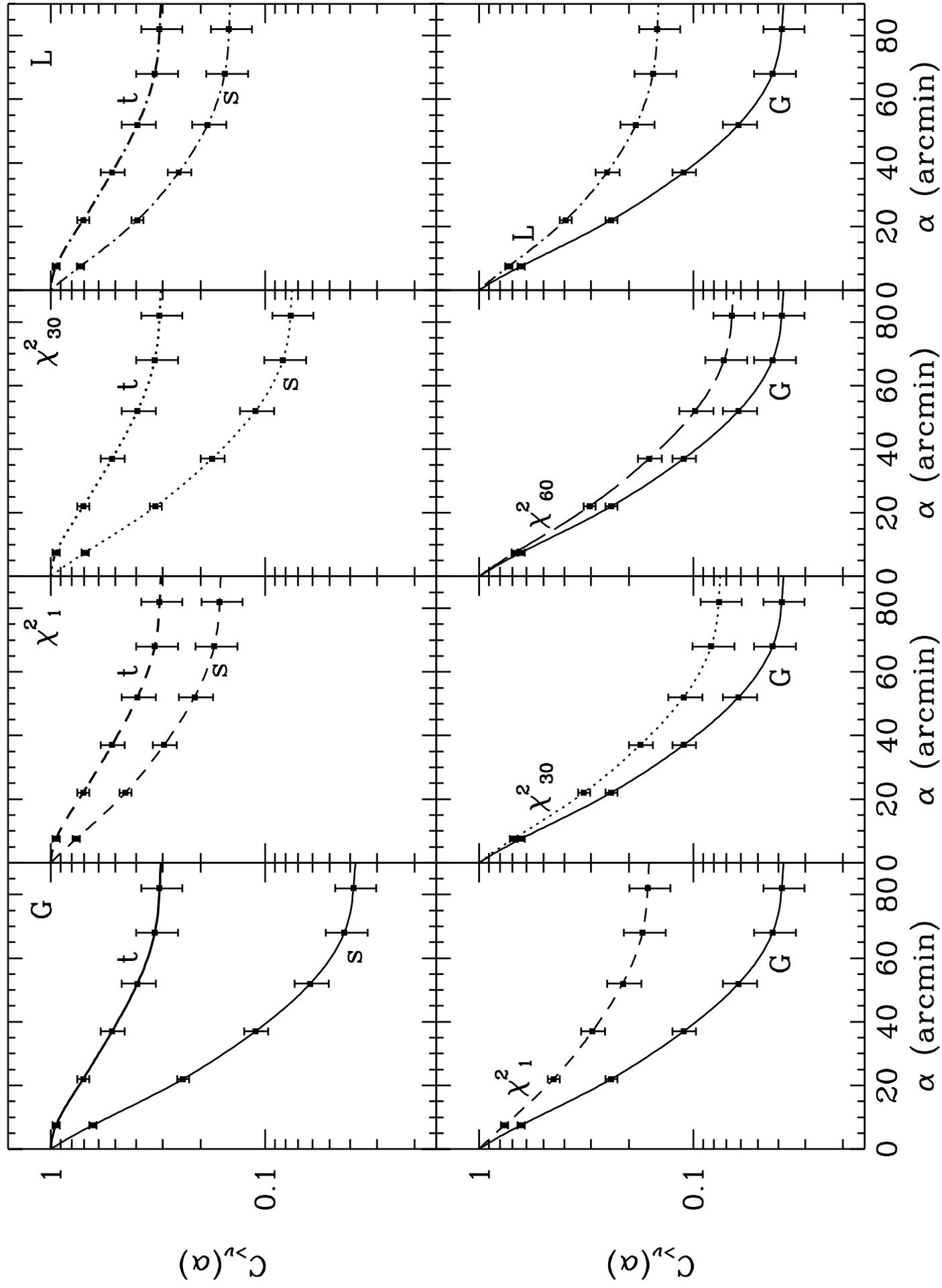
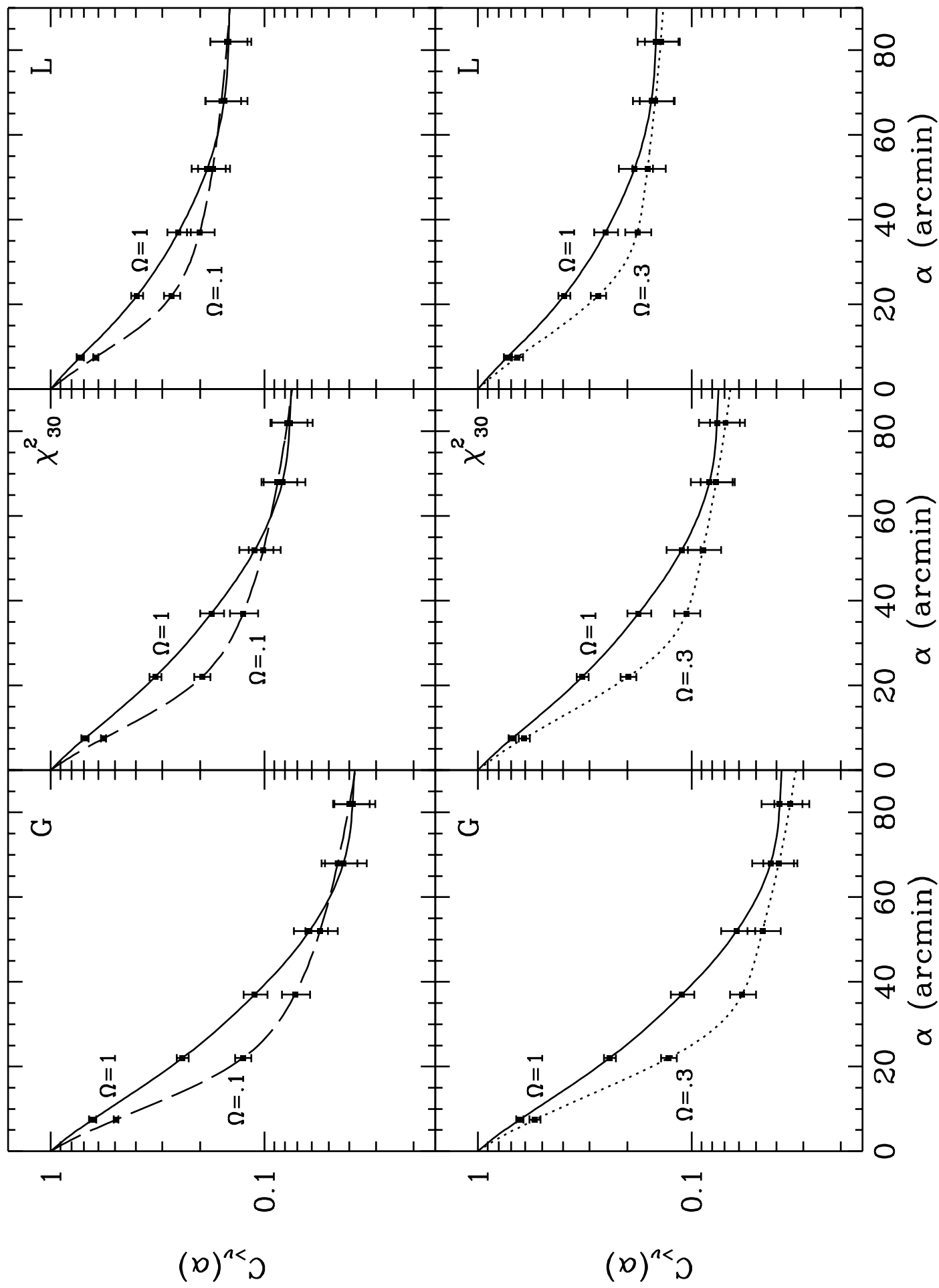
\end{document}